\documentclass{article}
\usepackage{graphics}
\usepackage{graphicx}
\begin{document}

\author{J.F. G\'omez$^{a}$, J.J. Rosales$^{b}$, J.J.
Bernal$^{a}$, V.I. Tkach$^c$,
M. Gu\'ia$^b$\\
\\
$^a$Departamento de F\'isica\\
Divisi\'on de Ciencias e Ingenier\'ias Campus Le\'on\\
Universidad de Guanajuato\\
Lomas del Bosque s/n, Lomas del Campestre\\
Le\'on Guanajuato. M\'exico\\
\\
$^b$Departamento de Ingenier\'ia El\'ectrica \\
Divisi\'on de Ingenier\'ias Campus Irapuato-Salamanca\\
Universidad de Guanajuato\\
Carretera Salamanca-Valle de Santiago, km. 3.5 + 1.8 km\\
Comunidad de Palo Blanco, Salamanca Guanajuato. M\'exico\\
\\
$^c$Department of Physics and Astronomy\\
Northwestern University\\
Evanston, IL 60208-3112, USA}
\title{Fractional Electromagnetic Waves}
\maketitle

{\bf Abstract:} In the present work we consider the
electromagnetic wave equation in terms of the fractional
derivative of the Caputo type. The order of the derivative being
considered is $0<\gamma\leq 1$. A new parameter $\sigma$ is
introduced which characterizes the existence of the fractional
components in the system. We analyze the fractional derivative
with respect to time and space, for $\gamma=1$ and $\gamma=1/2$ cases.\\


The recent interest on the fractional calculus (FC) and in
particular in the fractional differential equations is stimulated
by the applications in various areas of physics, chemistry,
engineering and bioengineering [1]-[5]. Nevertheless, the
derivation of such equations from some fundamental laws is not an
easy matter. The fractional operator reflects intrinsic
dissipative processes that are sufficiently complicated in nature.
Their theoretical relationship with FC is not yet fully
ascertained. Therefore, it is interesting to analyze a simple
physical system and try to understand their complete behavior
given by the fractional differential equation.
\\

In this work we will consider the electromagnetic wave equation in
terms of the fractional derivative of the Caputo type. The
solutions to the fractional differential wave equation are given
in terms of the Mittag-Leffler function. First, we consider the
fractional derivative with respect to time and second, the
fractional derivative with respect to space.
\\

The Maxwell equations for the electromagnetic waves in matter may
be written as
\begin{eqnarray}
\vec{\nabla}\cdot {\vec E} &=& \frac{4\pi}{\epsilon} \rho(\vec{r},t), \label{1}\\
\vec{\nabla}\cdot {\vec B} &=& 0, \label{2}\\
\vec{\nabla}\times {\vec E} &=& - \frac{1}{c} \frac{\partial \vec
B}{\partial t}, \label{3}\\
\vec{\nabla}\times{\vec B} &=& \frac{4\pi \mu}{c}{\vec
j(\vec{r},t)} + \frac{\epsilon \mu}{c}\frac{\partial {\vec
E}}{\partial t}, \label{4}
\end{eqnarray}
where $\rho(\vec{r},t)$ and $j(\vec{r},t)$ are general,
time-dependent distributions of charge densities and current
densities, respectively. In (\ref{4}) $\vec{D} = \epsilon \vec{E}$
is the dielectric displacement which is proportional to the
electric field $\vec{E}$ with the electric constant permittivity
$\epsilon$, and the magnetic field $\vec{B} = \mu \vec{H}$,where
$\vec{H}$ is the magnetic field intensity and $\mu$ the magnetic
permeability. In the presence of matter the magnetic field
intensity $\vec{H}$ replaces the magnetic induction vector
$\vec{B}$, in vacuum these field quantities are equal to each
other $\vec H = \vec B$. In the case of homogenous and isotropic
medium the parameters $\epsilon$ and $\mu$ are constants,
otherwise are vectors. Introducing the potentials, vector
$\vec{A}(x_i,t)$ and scalar $\phi(x_i,t)$
\begin{equation}
\vec{B} = \vec{\nabla} \times \vec{A}, \label{5}
\end{equation}
\begin{equation}
\vec{E} = -\frac{1}{c}\frac{\partial \vec{A}}{\partial t} -
\vec{\nabla} \phi, \label{6}
\end{equation}
and using the Lorenz gauge condition we obtain the following
decoupled differential equations for the potentials
\begin{eqnarray}
\Delta\vec{A}(\vec{r},t) - \frac{\epsilon
\mu}{c^2}\frac{\partial^2 \vec{A}(\vec r, t)}{\partial t^2}
&=&-\frac{4\pi}{c}\vec{j}(\vec
r, t), \label{7}\\
\Delta\phi(\vec{r},t) - \frac{\epsilon \mu}{c^2}\frac{\partial^2
\phi(\vec r, t)}{\partial t^2} &=&-\frac{4\pi}{\epsilon}
\vec{\rho}(\vec r, t),\label{8}
\end{eqnarray}
where $\frac{\epsilon \mu}{c^2}=\frac{1}{\upsilon^2}$. $\upsilon$
is the velocity of the light in the medium.\\

The idea is to write the ordinary differential wave equations
(\ref{1},\ref{2},\ref{3},\ref{4}) and (\ref{7},\ref{8}) in the
fractional form with respect to $t$. For this, we propose to
change the ordinary time derivative operator by the fractional in
the following way
\begin{equation}
\frac{d }{dt} \to \frac{1}{\sigma^{1 -
\gamma}}\frac{d^{\gamma}}{dt^{\gamma}}, \qquad n-1<\gamma\leq n,
\label{9}
\end{equation}
where $\gamma$ is an arbitrary parameter which represents the
order of the time derivative, $0 < \gamma \leq 1$, and, $\sigma$,
is a new parameter representing the fractional time components in
the system, its dimensionality is the second. In the case
$\gamma=1$ the expression (\ref{9}) transforms into ordinary time
derivative operator
\begin{equation}
\frac{1}{\sigma^{1 - \gamma}}\frac{d^{\gamma}}{
dt^{\gamma}}\Big|_{\gamma = 1} = \frac{d }{dt}. \label{10}
\end{equation}
The following Caputo definition of the fractional derivative will
be used [1],
\begin{eqnarray}
\frac{d^{\gamma}}{dt^{\gamma}} f(t) &=& \frac{1}{\Gamma(n -
\gamma)}\int_0^t\frac{f^{(n)}(\tau)}{(t-\tau)^{\gamma+1-n}}d\tau,
\label{11}\\
&& n-1< \gamma \leq n \in {I\!\!N} = \lbrace 1,2,...\rbrace,
\nonumber
\end{eqnarray}
where $\gamma \in {I\!\!R}$ is the order of the fractional
derivative
\begin{equation}
f^{(n)}(\tau) = \frac{d^n f(\tau)}{d\tau^n}, \label{11a}
\end{equation}
and
\begin{equation}
\Gamma(x) = \int_0^{\infty}e^{-t} t^{x-1} dt, \label{11b}
\end{equation}
is the gamma function.
\\

First, we consider the fractional time derivative. Then, using
(\ref{10}) the Maxwell equations (\ref{1}-\ref{4}) may be written
in terms of the fractional time derivatives
\begin{eqnarray}
\vec{\nabla} \cdot \vec{E} &=& \frac{4\pi \rho}{\epsilon},
\label{12}\\
\vec{\nabla} \cdot \vec{B} &=& 0, \label{13}\\
\vec{\nabla}\times \vec{E} &=& -\frac{1}{c} \frac{1}{\sigma^{1 -
\gamma}} \frac{\partial^{\gamma} \vec{B}}{\partial t^{\gamma}},
\label{14}\\
\vec{\nabla}\times \vec{B} &=& \frac{4\pi \mu}{c}\vec{j} +
\frac{\epsilon \mu}{c} \frac{1}{\sigma^{1-\gamma}}
\frac{d^{\gamma} \vec{E}}{\partial t^{\gamma}}.\label{15}
\end{eqnarray}
The relations (\ref{5},\ref{6}) become
\begin{equation}
\vec{B} = \vec{\nabla} \times \vec{A}, \label{5a}
\end{equation}
\begin{equation}
\vec{E} = -\frac{1}{c\sigma^{1-\gamma}}\frac{\partial^{\gamma}
\vec{A}}{\partial t^{\gamma}} - \vec{\nabla} \phi. \label{6a}
\end{equation}
Then, applying the Lorentz gauge condition we obtain the
corresponding time fractional wave equations for the potentials
(\ref{7},\ref{8})
\begin{eqnarray} \Delta\vec{A} - \frac{\epsilon \mu}{c^2}
\frac{1}{\sigma^{2(1 -\gamma)}} \frac{\partial^{2\gamma}}{\partial
t^{2\gamma}} \vec{A}
&=& - \frac{4\pi \mu}{c} \vec{j}, \label{16}\\
\Delta\phi - \frac{\epsilon \mu}{c^2} \frac{1}{\sigma^{2(1
-\gamma)}} \frac{\partial^{2\gamma}}{\partial t^{2\gamma}} \phi
&=& - \frac{4\pi}{\epsilon} \rho. \label{17}
\end{eqnarray}
In the case, $\gamma=1$, the equations (\ref{16}) and (\ref{17})
become (\ref{7}) and (\ref{8}).
\\

If, $\rho=0$, and, $\vec{j}=0$, we have the homogeneous fractional
differential equations
\begin{eqnarray}
\Delta\vec{A} - \frac{\epsilon \mu}{c^2} \frac{1}{\sigma^{2(1
-\gamma)}} \frac{\partial^{2\gamma}}{\partial t^{2\gamma}} \vec{A}
&=& 0, \label{18}\\
\Delta\phi - \frac{\epsilon \mu}{c^2} \frac{1}{\sigma^{2(1
-\gamma)}} \frac{\partial^{2\gamma}}{\partial t^{2\gamma}} \phi
&=& 0. \label{19}
\end{eqnarray}
We are interested in the analysis of the electromagnetic fields in
the medium starting from the equations (\ref{18}) and (\ref{19}).
We can write the fractional equations (\ref{18}) and (\ref{19}) in
the following compact form
\begin{equation}
\frac{\partial^2 z(x,t)}{\partial x^2} - \frac{\epsilon
\mu}{c^2}\frac{1}{\sigma^{2(1-\gamma)}} \frac{\partial^{2\gamma}
z(x,t)}{\partial t^{2\gamma}} = 0, \label{20}
\end{equation}
where $z(x,t)$ represents both $\vec{A}(x,t)$ and $\phi(x,t)$. We
consider a polarized electromagnetic wave, then $A_x(x_i,t) = 0$
$A_y(x_i,t) \neq 0$ $A_z(x_i,t) \neq 0$. The equation (\ref{20})
is lineal and a particular solution may be found in the form
\begin{equation}
z(x,t) = z_0 e^{-ikx}\cdot u(t), \label{21}
\end{equation}
where $k$ is the wavevector in the $x$ direction and $z_0$ is a
constant. Substituting (\ref{21}) into (\ref{20}), we obtain
\begin{equation}
\frac{\partial^{2\gamma} u(t)}{\partial t^{2\gamma}} +\upsilon^2
k^2 \sigma^{2(1-\gamma)}u(t)=0.\label{22}
\end{equation}
Redefining
\begin{equation}
\omega^2 =\upsilon^2 k^2 \sigma^{2(1-\gamma)}=\omega_{0}^{2}
\sigma^{2(1-\gamma)}, \label{23}
\end{equation}
where $\omega_{0}$ is the fundamental frequency of the
electromagnetic wave, the equation (\ref{22}) may be written as
\begin{equation}
\frac{\partial^{2\gamma} u(t)}{\partial t^{2\gamma}} + \omega^2
u(t) = 0.\label{24}
\end{equation}
The solution of this equation may be found in the form of the
power series. The solution is
\begin{equation}
u(t) = E_{2\gamma}\Big(-\omega^2 t^{2\gamma}\Big), \label{25}
\end{equation}
where
\begin{equation}
E_{2\gamma}(-\omega^2 t^{2\gamma}) = \sum_{n=0}^{\infty}
\frac{(-\omega^2 t^{2\gamma})^n}{\Gamma(2n\gamma + 1)}, \label{26}
\end{equation}
is the Mittag-Leffler function. Substituting the expression
(\ref{25}) in (\ref{21}) we have a particular solution of the
equation (\ref{20})
\begin{equation}
z(x,t) = z_0 e^{-ikx}\cdot E_{2\gamma}\Big(-\omega^2
t^{2\gamma}\Big). \label{21a}
\end{equation}

{\bf In the first case}, $\gamma = 1$, the Mittag-Leffler function
(\ref{25}) transforms into hyperbolic cosines and, from (\ref{23})
$\omega = \omega_0$. Then
\begin{equation}
E_2(-\omega_{0}^{2} t^2)= {\rm cosh}{\Big(\sqrt{-\omega_{0}^{2}
t^2}\,\,\Big)} = {\rm cosh}(i\omega_0 t) = {\rm cos}(\omega_{0}t).
\label{27}
\end{equation}
The expression (\ref{27}) is a periodic function with respect to
$t$. Therefore, in the case $\gamma = 1$, the solution to the
equation (\ref{20}) is
\begin{equation}
z(x,t) = {\rm Re}z_0 e^{i(\omega_0t - kx)},
\label{28}
\end{equation}
which defines a periodic, with fundamental period $T_0 =
\frac{2\pi}{\omega_0}$, monochromatic wave in the, $x$, direction
and in time, $t$. This
result is very well known from the ordinary electromagnetic waves theory.\\

{\bf For the second case}, $\gamma = 1/2$, the equation (\ref{20})
becomes
\begin{equation}
\frac{\partial^2 z(x,t)}{\partial x^2} - \frac{\epsilon \mu}{c^2}
\frac{1}{\sigma} \frac{\partial z(x,t)}{\partial t} = 0.\label{29}
\end{equation}
The solution may be found in the form of ({\ref{21}), then we
obtain the following equation for the function $u(t)$
\begin{equation}
\frac{du}{dt} + \omega^2 u(t) = 0, \label{30}
\end{equation}
where, in this case, $\omega^2 = \omega_0^2 \sigma$, from
(\ref{23}). Solution of the equation (\ref{30}) may be obtained in
terms of the Mittag-Leffler function (\ref{25}). In the case,
$\gamma = 1/2$, we have
\begin{equation}
u(t) = E_1\lbrace -\omega^2 t \rbrace = e^{-\omega^2 t}.
\label{31}
\end{equation}
The particular solution is
\begin{equation}
z(x,t) = z_0 e^{-\omega^2 t}e^{-ikx},  \label{32}
\end{equation}
For this case the solution is periodic only respect to $x$ and it
is not periodic with respect to $t$. The solution represents a
plane wave with time decaying amplitude. The time in which the
amplitude $z_0$ decay $e$ times is
\begin{equation}
t_0 = \frac{1}{\omega^2} = \frac{1}{\omega_0^2\sigma}. \label{33}
\end{equation}
It is important to note that, $\gamma$, is a dimensionless
quantity which characterizes the order of fractional time
derivative while the quantity, $\sigma$, has dimensions of time,
and characterizes the presence of fractional time components in
the medium. However, these two quantities are related as follows
\begin{equation}
\gamma=\sigma^2
\omega_{0}^{2}=\frac{\sigma^2}{T_{0}^{2}}=\frac{\sigma_{x}^{2}}{\lambda^2},
\qquad 0<\sigma\leq T_{0}. \label{34}
\end{equation}
where, $T_{0}$, is the period of the wave, $\lambda$, is the
wavelength and, $\sigma=\frac{\sigma_{x}}{\upsilon}$, where, $v$,
is the velocity of the electromagnetic wave in the,
$x$, direction.\\

Taking into account this relation, the solution (\ref{25}) may be
written as
\begin{equation}
u(t)=E_{2\gamma}\Big(-\gamma^{(1-\gamma)}\tilde{t}^{2\gamma}\Big),
\label{35}
\end{equation}
where, $\tilde{t}=\frac{t}{T_{0}}$, is a dimensionless parameter.\\

Now, we will consider the equation (\ref{20}) assuming that the
spatial derivative is fractional and the time derivative is
ordinary. Then, we have the spatial fractional equation
\begin{equation}
\frac{1}{\sigma_x^{2(1 - \delta)}}\frac{\partial^{2\delta}
\tilde{z}(x,t)}{\partial x^{2\delta}}
-\frac{1}{\upsilon^2}\frac{\partial^2 \tilde{z}(x,t)}{\partial
t^2} = 0, \label{36}
\end{equation}
where the order of the fractional differential equation is
represented by $0<\delta\leq 1$, and $\sigma_x$ has length
dimension. A particular solution to the equation (\ref{36}) may be
as follows
\begin{equation}
\tilde{z}(x,t) = \tilde{z_0} e^{i\omega t} u(x). \label{37}
\end{equation}
Substituting (\ref{37}) in (\ref{36}), we obtain
\begin{equation}
\frac{\partial^{2\delta}u(x)}{\partial {x}^{2\delta}}+\tilde{k}^2
u(x)=0, \label{38}
\end{equation}
where
\begin{equation}
\tilde{k}^2=\frac{\omega^2}{\upsilon^2}\sigma_{x}^{2(1-\delta)} =
k^2\sigma_{x}^{2(1-\delta)}, \label{38a}
\end{equation}
is the wave-vector in the medium in presence of fractional
components, and $k$ is the wave vector in the medium without its
presence. The wave-vectors are equal, $\tilde k = k$, only in the
case, $\delta = 1$. Solution of the equation (\ref{38}) is given
in terms of the Mittag-Leffler function
\begin{equation}
u(x)=E_{2\delta}(-\tilde{k}^2
x^{2\delta})=\sum_{n=0}^{\infty}\frac{(-\tilde{k}^2x^{2\delta})^n}{\Gamma(2n\delta+1)}.\label{39}
\end{equation}
{\bf First case:} For the fractional spatial case, when $\delta =
1$, from (\ref{38a}) and (\ref{39}), we have
\begin{equation}
E_{2}(-k^2x^2) = {\rm cosh}(\sqrt{-k^2x^2}) = {\rm cosh}(-ikx) =
{\rm Re}(e^{-ikx}).\label{40}
\end{equation}
In this case the solution follows from (\ref{37})
\begin{equation}
\tilde{z}(x,t) = {\rm Re}{\tilde{z_0}}e^{i\omega t - i{k}x},
\label{41}
\end{equation}
with, $\tilde{k} = k = \frac{\omega}{\upsilon}$, where, $k$, is
the component of the wave-vector in the, $x$, direction and is
related with the wavelength by, $k = \frac{1}{\lambda}$. The solution (\ref{41}) represents
a periodic, with respect to $t$ and $x$, monochromatic wave. \\

{\bf Second case:} For the case $\delta = 1/2$, we have from
(\ref{38a}), $\tilde{k^2} = k^2 \sigma_x =
\frac{\omega^2}{\upsilon^2}\sigma$, and $[\tilde{k^2}] =
\frac{1}{l}$ has dimensions of the inverse of the length. The
solution for this case, has the form
\begin{equation}
u(x) = E_1\Big(-\tilde{k^2}x \Big) = e^{-\tilde{k^2}x}. \label{42}
\end{equation}
The solution (\ref{37}) is written as
\begin{equation}
\tilde{z}(x,t) = {\tilde{z_0}}e^{i\omega t}e^{-\tilde{k}^2x}.
\label{43}
\end{equation}
The wave is periodic only with respect to $t$. The distance at
which the amplitude $\tilde{z}_{0}$ is reduced $e$ times is
\begin{equation}
x_0 = \frac{1}{\tilde{k^2}} = \frac{1}{k^2 \sigma_x}. \label{44}
\end{equation}
In this case we have that, $\delta$, is a dimensionless quantity
and $\sigma_{x}$ is related to the fractional space. These two
quantities are related by
\begin{equation}
\delta = k^2\sigma_{x}^{2}=\frac{\sigma_{x}^{2}}{\lambda^2}.
\label{45}
\end{equation}
We can use this relation in order to write the equation (\ref{39})
as follows \begin{equation} u(x) =
E_{2\delta}\Big(-\delta^{(1-\delta)}\tilde{x}^{2\delta}\Big),
\label{46}
\end{equation}
where, $\tilde{x}=\frac{x}{\lambda}$, is a dimensionless
parameter. It can be seen that the solutions (\ref{35}) and
(\ref{46}) have the same structure, and in the case,
$\delta=\gamma$, have the same shape. Then, we can plot the
function $u(s)$ where $s =(\tilde{x}, \tilde{t})$, for different
values of the fractional parameter $\gamma$, (see Figure 1).
\\

\begin{figure}[h]
\begin{center}
\includegraphics[width=9cm]{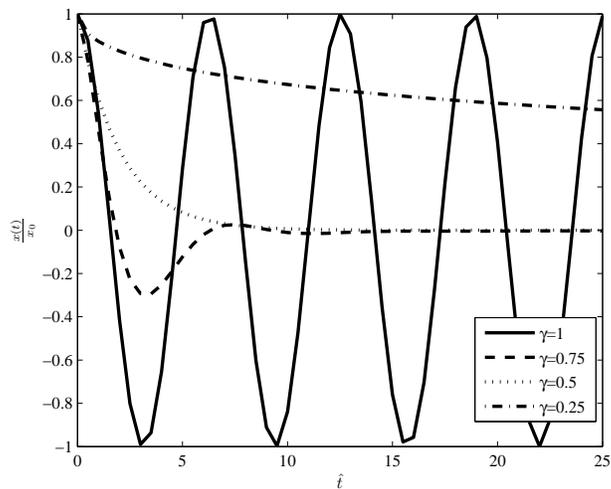}
\caption{Graph corresponding to the equations (\ref{35}) and
(\ref{46})}.
\end{center}
\end{figure}

{\bf Conclusion:} In this work we have studied the behavior of the
electromagnetic waves applying the formalism of the fractional
calculus. The order of the derivative being considered is
$0<\gamma\leq 1$. It showed that for the case where $\gamma =
\delta = 1$ the solutions represent a periodic, with respect to
$t$ and $x$, monochromatic wave, as it should be. However, if we
take $\gamma = 1/2$ the periodicity with respect to $t$ is broken
and behaves like a wave with time decaying amplitude
Eq.(\ref{32}). On the other hand, when $\delta = 1/2$ the
periodicity with respect to $x$ is broken and behaves like a wave
with spatial decaying amplitude Eq.(\ref{43}).

\end{document}